\newcommand{\beq}{\begin{equation}}
\newcommand{\eeq}{\end{equation}}
\newcommand{\beeq}{\begin{eqnarray}}
\newcommand{\eeeq}{\end{eqnarray}}

\documentstyle[12pt,epsfig] {article}

\begin{document}

\begin{center}\bf
COMPOSITE SUPERSTRING MODEL FOR HADRONS AND EXTENDED VIRASORO
SUPERCONFORMAL SYMMETRY WITH SUPERCURRENTS CONSTRAINTS.

\bigskip
V. A.  Kudryavtsev  \\                                 Petersburg
 Nuclear Physics Institute\\
\bigskip
A b s t r a c t
\end{center}

 Hadron dynamics is formulated in terms of interacting composite strings.
 Consistent composite string model with extended Virasoro superconformal
 symmetry is found. These composite strings carry flavour and chiral
 degrees of freedom on edging two-dimensional surfaces. Necessary correct
 description of string amplitudes without states of negative norms in the
 spectrum of physical states is reached when special supercurrents conditions
 are fulfiled.

\newpage

\section{Introduction. Some problems of string models for hadrons. }

     An essential interest in string description of hadron interactions
 has  arised as far back as forty years ago (the Nambu string [1] and dual
resonance models initiated by Veneziano's work [2])   due to the
remarkable universal linearity of Regge trajectories $\alpha(t)$ for
meson and baryon resonances [3]
$J=\alpha(M^2)=\alpha(0)+\alpha^\prime_H M^2$
$(\alpha^\prime_H\approx0.85\,$GeV$^{-2})$;  where J,M are spin and
mass of a resonance. Now we have these trajectories up to $J=5$ and
states for not only leading (n=0) but for second (n=1),for third
(n=2)and even for fourth (n=3)  daughter trajectories
$J_n=\alpha(0)-n+\alpha^\prime_HM^2$ n=0,1,2,3... . See [4] .

     However attempts to build the string model for hadron interactions have
not been succesful since consistent models for relativistic quantum
strings [5]   have required the intercept of leading meson
trajectory $\alpha(0)$ to be equal to one. A shift of this value
from one has led to contradiction  with unitarity. The real leading
$\rho$-meson trajectory has the intercept to be equal one half
approximately. Just this reason has led to superstring models  with
massless gluons (open strings) on the trajectory
$J=1+\alpha^\prime_PM^2$ and for masslees gravitons (closed strings)
on the trajectory$J=2+1/2\alpha^\prime_PM^2$.

     Here we are facing  other
problem for hadron dynamics in the framework of ordinary string
description. In the consistent critical case of string models
nonplanar loop string diagrams  lead to appearance of closed string
states as bound states of open strings (Fig.1). Then we shall obtain
the slope $\alpha^\prime_{closed}$ for closed strings as half of
$\alpha^\prime_{open}$ for open strings following usual string
models. But a natural scale parameter $\alpha^\prime$ for the
graviton trajectory  would be corresponding to the Planck mass
$10^{19}$ Gev   i.e.
$(\alpha^\prime_{closed}\equiv\alpha^\prime_P\sim10^{-38}$GeV$^{-2})$.
Certainly it is beyond reach of hadron interactions scale
corresponding to $\alpha^\prime_H\approx0,85$GeV$^{-2}$.
   As we shall see both problems find a solution in the context
of composite string model.

\begin{figure}
\begin{center}
\centerline{\epsfig{figure=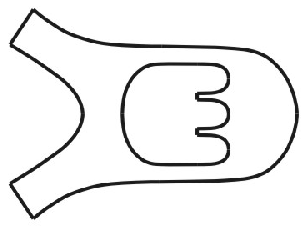,width=0.50\textwidth,clip=}
            \hspace{0.3cm}
 \epsfig{figure=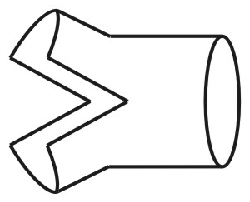,width=0.50\textwidth,clip=}}
Fig.1.
\end{center}
\end{figure}

\section{New class of relativistic quantum strings: composite strings }

       Traditional consistent models for open strings have N=1
   superconformal Virasoro symmetry on two-dimensional world
   surface. This superconformal Virasoro algebra leads us in critical
   case to the intercept of leading boson trajectory to be equal to
   one and to massless vector bosons in the spectrum of physical
   states correspondingly. For closed strings as bound states of two
   open strings we have a pair of two-dimensional surfaces and the
   N=2 superconformal Virasoro symmetry. These quantum
   superconformal symmetries are a reflection of classical conformal
   symmetries on two-dimensional  world surfaces for string actions [6] .

     A generalization of classical multireggeon (multistring)
 vertices [7] has been suggested by author in 1993 [8].
   These string amplitudes have been used for description of
   interaction of many $\pi$-mesons [9].
   New string vertices give a new geometric picture for interactions
   of strings which has a natural description in terms of
   composite strings and three two-dimensional surfaces for moving
   open string instead of usual one (see Fig.2.).

\begin{figure}
\begin{center}
\centerline{\epsfig{figure=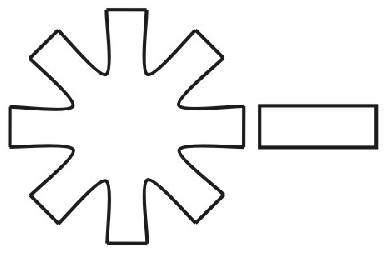,width=0.80\textwidth,clip=}
            \hspace{0.3cm}
 \epsfig{figure=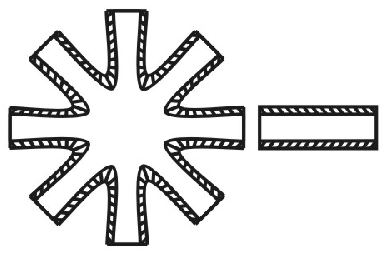,width=0.80\textwidth,clip=}}
Fig.2.
\end{center}
\end{figure}

    These additional edging two-dimensional surfaces carry quark
    quantum numbers (flavour, spin, chirality). This composite
    string  construction reminds a gluon string with two pointlike
    quarks at ends of this string  or a simplest case of a string ending at two
    branes     when they are themselves some strings (Fig.3).

\begin{figure}
\begin{center}
\centerline{\epsfig{figure=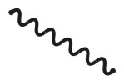,width=0.50\textwidth,clip=}
\hspace{0.3cm}
 \epsfig{figure=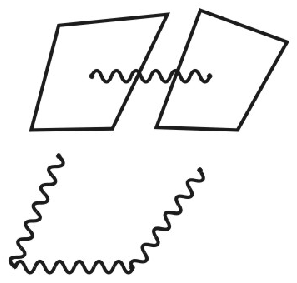,width=0.50\textwidth,clip=}}
Fig.3.
\end{center}
\end{figure}

     It is not surprisingly as we shall see further that we have here
     a possibility for N=3 extended superconformal Virasoro symmetry
     for these composite strings. Let us note that we have no supersymmetry
     in the Minkovsky (target) space for this model. The topology of interacting
     composite strings allows to solve the problem of the intercept
     $\alpha(0)$ for leading meson trajectory and to shift  the
     value of this intercept to one half without breakdown of the
     extended superconformal Virasoro symmetry for composite strings
     due to non-vanishing conformal weights for fields on both edging
     two-dimensional surfaces.

          Composite origin of objects under consideration brings to
     solution of second problem for hadron strings. Indeed there is
     a new parameter for composite open strings in addition to usual
     parameter $\alpha^\prime$ for classical open strings. This new
     parameter defines the part of momentum of hadron composite
     string which flows in central (basic) two-dimensional surface
     in reference to the rest of momentum on edging surfaces. If
     this part is vanishing then unitarity breaks down due to appearance
     of some nonunitary fixed singularity without any massless tensor state
      (graviton) for nonplanar loop diagrams.
     The non-vanishing value of this ratio defines the value of other
     ratio $\alpha^\prime_{closed}$/$\alpha^\prime_{open}$.
     It follows from nonplanar one-loop diagram for composite
     strings which lead to states of closed strings without edging surfaces(Fig.4).

         \begin{figure}
\begin{center}
\mbox {\epsfig{file=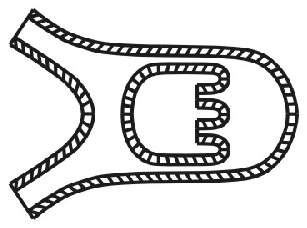,width=0.70\textwidth}}\\
  Fig.4.
\end{center}
\end{figure}

         So we have a possibility to choose  this ratio as it is
      necessary for the correct graviton pole as
      $\alpha^\prime_{closed}$/$\alpha^\prime_{open}$$\sim10^{-38}$.

        This circumstance solves one more problem of asymptotic
        behaviour of hadron string amplitudes in deep inelastic
        region. Here we have Regge trajectories with the very small
        slope for closed string sector. Such Regge trajectories lead
        us to approximately power behaviour in deep inelastic region
        and at high energies s for fixed ratio s/t.

\section{Formulation of composite string model and vertices for interacting composite strings.}

       Many-string vertices of interacting composite open strings  are
natural generalizations of corresponding many-string
       (many-reggeon) vertices for classical string models.
        The classical Lovelace-Olive-Alessandrini multi-string vertices [7]
 define an interaction of $N$ arbitrary string states in the
Veneziano or the Neveu--Schwarz-Ramond models [6] in terms of
two-dimensional string fields:  $X^{(i)}_\mu(z_i)$ (the coordinate
of $i$-th string) for the vanishing conformal spin j and
$H^{(i)}_\mu(z_i)$ (the anticommuting superpartner of
$X^{(i)}_\mu(z_i)$) for the conformal spin j to be equal to
$\frac12$:
$$
X^{(i)}_\mu(z_i)\ =\ X^{(i)}_{o\mu}+P^{(i)}_\mu\ln{z_i} +\sum_n
\frac{a^{(i)}_{n\mu}}{in}\,z^n\ ,
$$
 $a^+_{n\mu}=a_{-n\mu}$ ,   \\
$a_n|0\rangle=\langle0|a_{-n}=0$  ;\  $n>0$\ ,
$$
\left[ a^{(i)}_{n\mu},a^{(i)}_{m\nu}\right]\ =\ -ng_{\mu\nu}
\delta_{n,-m}\ ,
$$
$$
H^{(i)}_\mu(z_i)=\sum b^{(i)}_{r\mu}z^r\ , \qquad \left\{
b^{(i)}_{r\mu}, b^{(i)}_{s\nu}\right\}= -g_{\mu\nu}\delta_{r,-s}\ ,
$$
$b_r^+= b_{-r}$ $b_r|0\rangle\ =\ \langle0|b_{-r}=0$ .

The $N$-string amplitudes are represented by some integrals over
$z_i$ variables of the vacuum expectation value for a product of a
vertex operator and wave functions of string states:
\begin{equation}
  A_N = \int\prod dz_i \langle0|V_n \prod_i\hat{\Psi^i}|0\rangle\ ,
\end{equation}
where $\hat{\Psi^i}|0\rangle$ is a wave function of $i$-th string
state.

The $N$-string vertex operator $V_N$ for the Neveu--Schwarz model is
given by the following exponent:
\begin{eqnarray}
V_N^{NS}  &=& \exp\bigg(\frac 12\sum_{n,m,p \atop i \neq j}
\frac{a^{(i)}_{n}}{\sqrt{n}}
(U^{(i)}_\varepsilon)_{nm}(V^{(j)}_\varepsilon)_{mp}
\frac{a^{(j)}_{p}}{\sqrt{p}}\ + \nonumber\\
&+& \frac 12 \sum_{n,m,p \atop i \neq
j}b^{(i)}_{n+1/2}(U^{(i)}_{1/2})_{nm}
(V^{(j)}_{1/2})_{mp}b^{(j)}_{p+1/2}\bigg)\ ,
\end{eqnarray}
where $(U^{(i)}_j)_{nm}$, $(V^{(i)}_\j)_{mp}$
  are the special infinite matrices
which depend on $z_{i-1}, z_i, z_{i+1}$ complex variables .They are some representations of the SU(1,1)group
for the conformal spin j.\\
     These vertices $V_N^{NS}$ have the necessary factorization and
conformal properties.

       It turns out there is another operator $W_N$ with these properties:
\begin{eqnarray}
W_N &=&
\sum_{n,m,k}\widetilde\Psi^{(1)}_{n+\frac12}(U^{(1)}_{\frac12})_{nm}
(V^{(2)}_{\frac 12})_{mk}\Psi^{(2)}_{k+\frac
12}\sum_{l,p,s}\widetilde \Psi^{(2)}_{l+\frac 12} (U^{(2)}_{\frac
12})_{lp}(V^{(3)}_{\frac 12})_{ps} \Psi^{(3)}_{s+\frac 12} \ldots
\times \nonumber  \\& & \times  \sum  \widetilde \Psi^{(N)}
U^{(N)}_{\frac 12}V^{(1)}_{\frac12} \Psi^{(1)}\equiv \\& &
\prod^N_{i=1}\sum_{n,m,p} \widetilde \Psi^{(i)}_{n+\frac
12}(U^{(i)}_{\frac 12})_{nm}(V^{(i+1)}_{\frac 12})_{mp}
\Psi^{(i+1)}_{p+\frac 12}    \nonumber .
\end{eqnarray}
$W_N$ is some cyclic symmetrical trace-like operator built from
Fourier components of the two-dimensional anticommuting fields
$\Psi^{(i)}(z)$. These new fields are similar to the Neveu-Schwarz
fields $H^{(i)}$ and have the conformal spin $j=\frac12$.
$$
\Psi^{(i)}_\alpha(z_i)\ =\ \sum\Psi^{(i)}_{r\alpha} z_i^r\ .
$$
 The generalized $N$-string vertex operator for composite strings
is the product of the old $V_N^{NS}$ and the new operator $W_N$
\begin{eqnarray}
 V_N^{comp}\ =\  V_N^{NS} W_N\ .
\end{eqnarray}

It is evident that the operators $V_N^{NS}$ and $W_N$  have the
different structures. The matrices $U^{(i)}$ and $V^{(j)}$ in
$V_N^{NS}$ connect all possible fields $X^{(i)}(a^{(i)}_n)$ with
each other. In the operator $W_N$ the matrices $U,V$ connect only
neighboring fields $\Psi^{(i)}$ and $\Psi^{(i+1)}$. That is why this
operator $W_N$ reproduces  the structure of dual quark diagrams here
and the operator $V_N^{NS} W_N$ leads us to composite objects i.e.
composite strings.

    It is possible to use other similar operators instead of (3) with
 $Y^{(i)}$-fields of j=1  and $f^{(i)}$-fields of $j= \frac{1}2$
  in this cyclic operator.
\newpage
    Namely
\begin{eqnarray}
 & &  W_N \tilde\psi^{(1)}\exp\bigg( \sum_{n,m,p}
\frac{Y^{(1)}_{n}}{\sqrt{n}}
(U^{(1)}_\varepsilon)_{nm}(V^{(2)}_\varepsilon)_{mp}
\frac{Y^{(2)}_{p}}{\sqrt{p}} +  \sum_{n,m,p}
f^{(1)}_{n+1/2}(U^{(1)}_{1/2})_{nm}
(V^{(2)}_{1/2})_{mp}f^{(2)}_{p+1/2}\bigg)\psi^{(2)} \nonumber \\
 & & \tilde \psi^{(2)}\exp\bigg(\sum_{n,m,p\atop}
\frac{Y^{(2)}_{n}}{\sqrt{n}}
(U^{(2)}_\varepsilon)_{nm}(V^{(3)}_\varepsilon)_{mp}
\frac{Y^{(3)}_{p}}{\sqrt{p}} +  \sum_{n,m,p}
f^{(2)}_{n+1/2}(U^{(2)}_{1/2})_{nm}
(V^{(3)}_{1/2})_{mp}f^{(3)}_{p+1/2}\bigg)\psi^{(3)}\ldots
\nonumber\\  & &  \tilde \psi^{(N)}\exp\bigg(\sum_{n,m,p}
\frac{Y^{(N)}_{n}}{\sqrt{n}}
(U^{(N)}_\varepsilon)_{nm}(V^{(1)}_\varepsilon)_{mp}
\frac{Y^{(1)}_{p}}{\sqrt{p}} +  \sum_{n,m,p}
f^{(N)}_{n+1/2}(U^{(N)}_{1/2})_{nm}
(V^{(1)}_{1/2})_{mp}f^{(1)}_{p+1/2}\bigg) \psi^{(1)}\nonumber   \\ &
& \equiv \prod^N_{i=1}\tilde \Psi^{(i)}\exp\bigg(\sum_{n,m,p}
\frac{Y^{(i)}_{n}}{\sqrt{n}}
(U^{(i)}_\varepsilon)_{nm}(V^{(i+1)}_\varepsilon)_{mp}
\frac{Y^{(i+1)}_{p}}{\sqrt{p}} + \\  & &   \sum_{n,m,p}
f^{(i)}_{n+1/2}(U^{(i)}_{1/2})_{nm}
(V^{(i+1)}_{1/2})_{mp}f^{(i+1)}_{p+1/2}\bigg) \psi^{(i+1)}.\nonumber
\end{eqnarray}

    For the composite string model under consideration we shall use the operator
     $V_N^{comp} =  V_N^{NS}W_N $ with $W_N$ of type (5). So far as just
     this  structure  has a symmetrical description of two-dimensional fields both on the basic and on the edging
     surfaces.
     It provides the necessary extended superconformal symmetry of composite string
     amplitudes as we shall see further.

    For investigation of composite superstrings it is more appropriate to move
from multi-string vertices (4) to more simple vertices $\hat{V}_i$
corresponding to emission of ground states. An amplitude $A_N$ of
interaction of N ground string states is represented by integral of
vacuum expectation of product of $\hat{V}_i$ vertices (Fig.5.)
\begin{eqnarray}
A_N = \int\prod dz_i\langle0|\hat{V}_1(z_1)\hat{V}_2(z_2)
\hat{V}_3(z_3)...\hat{V}_{N-1}(z_{N-1})\hat{V}_N(z_N)|0\rangle\,\nonumber\\
\hat{V}_i(z_i)\ =\ z^{-L_0}_i \hat{V}_i(1) z^{L_0}_i,\nonumber\\
\end{eqnarray}

\begin{figure}
\begin{center}
\centerline{\epsfig{figure=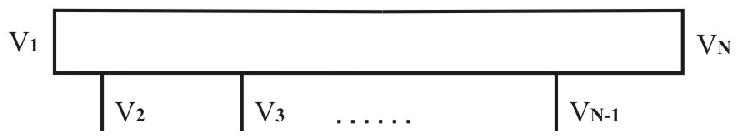,width=0.50\textwidth,clip=}}
Fig.5.
\end{center}
\end{figure}

  These vertices $\hat{V}_i$ have the well-known expressions for the
  Neveu-Schwarz model:

\begin{eqnarray}
&& \hat{V}_i(z_i)\ =\ z_i^{-L_0}\left[G_r,:\exp{ip_i X(1)}:\
\right]z_i^{L_0}\ ,\nonumber\\
&& :\exp{(i p_{i}X(1))} :\ =\ \exp {(i p_i X^{(+)}(1))}\ \exp{(ip_i
X_0 )} \exp{(ip_i X^{(-)}(1))}\
\end{eqnarray}
\begin{eqnarray}
G_r^{NS} &=& \frac1{2\pi}\int\limits_0^{2\pi}d\tau\left(H^{\mu}\frac
d{d\tau} X_\mu+ \hat{P}_\nu H^\nu\right)e^{-ir\tau}
\end{eqnarray}

\begin{eqnarray}
&& \hat{V}_i(1)= (p_i H(1)):\exp{(ip_i X(1))}\ :\equiv\,\nonumber\\
&& (\sum_r p_ib_r)\exp{(ip_i X^{(+)}(1))}\ \exp{(ip_i X_0)}\ \exp{(
ip_i X^{(-)}(1))}\ \equiv\,\nonumber\\
&& (\sum_r p_ib_r)\exp{(-ip_i\sum_n \frac {a_{-n}}{n})}\ \exp{(ip_i
X_0 )}\  \exp{(ip_i \sum_n \frac {a_{n}}{n})}\
\end{eqnarray}

      For composite strings operator structures (5) correspond to
 two-dimensional fields on edging surfaces.  The corresponding
 operator vertices $\hat{V}_i$ will contain as in (5) additional edging
 fields: $Y_{mu}$   $\mu$ = 0,1,2,3 and its superpartner $f_{\mu}$
  with Lorentz indices. In addition to them we include similar fields
 (J with superpartner $\Phi$) which carry internal quantum numbers
 (isospin and other flavours) on edging surfaces and corresponding I,$\Theta$ fields on the basic
 two-dimensional surface.  As we shall see some relations
 between fields on the basic (central) surface  and fields on edging two-dimensional
 surfaces ($\partial X_\mu ,H_{\mu}$ and $Y_{\mu}$,$f_{\mu}$ for
 Lorentz indices;   I,$\Theta$ and J,$\Phi$  for internal numbers)
 play an important role for the afore-mentioned extended Virasoro superconformal symmetry of the
 composite superstring model under consideration.

    Since the edging fields are propagating only on the  corresponding
 surfaces it is convenient to introduce vacuum states for the fields on the separate edging surfaces and
 to write (6) in equivalent form with help of these vacuum states:

\begin{eqnarray}
&& A_N =
 \int\prod dz_i\langle0^{(1,2)}|\hat{V}_{12}(z_1)\langle0^{(3)}|\hat{V}_{23}(z_2)\langle0^{(4)}||0^{(2)}\rangle\
\hat{V}_{34}(z_3) \langle0^{(5)}|...\\&& |0^{(i-1)}\rangle\
\hat{V}_{i,i+1}(z_i)\langle0^{(i+2)}|... |0^{(N-2)}\rangle\
\hat{V}_{N-1,N}(z_{N-1})|0^{(N-1)}\rangle\
\hat{V}_{N,1}(z_N)|0^{(N,1)}\rangle\ \nonumber
\end{eqnarray}

  This form(10) excludes this amplitude from the set of additive string models of the Lovelace's paper [5]
and leads  to the topology of composite string models [8,9]. Now we
are ready to formulate the vertex operator $\hat{V}_{i,i+1}(z_i)$
(Fig.6) for this composite string model:
\begin{eqnarray}
&& \hat{V}_{i,i+1}(z_i)\ =\ z_i^{-L_0}\left[G_r,\hat W_{i,i+1}
\right]z_i^{L_0}\ ,\nonumber\\
&&\hat W_{i,i+1}=\hat{R}^{out}_{i}\hat{R}_{NS}\hat{R}^{in}_{i+1}
\end{eqnarray}

 \begin{figure}
\begin{center}
\centerline{\epsfig{figure=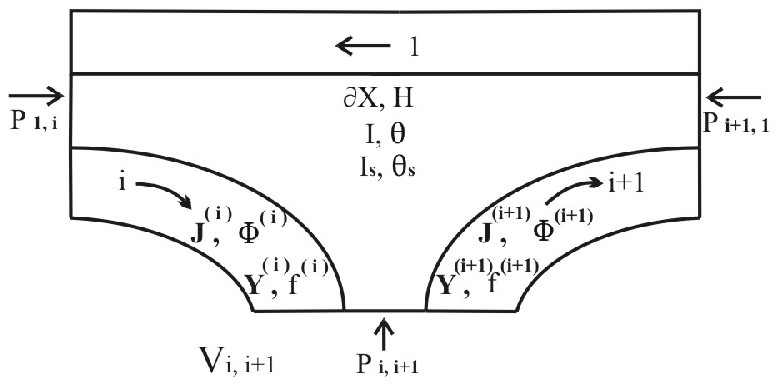,width=0.60\textwidth,clip=}}
Fig.6.
\end{center}
\end{figure}

 The operators $\hat{R}^{out}_{i}$and $\hat{R}^{in}_{i+1}$
are defined by fields on i-th and (i+1)-th  edging surfaces. The
operator $\hat{R}_{NS}$ is defined by fields on the basic surface.
They have the same structure as the operator $:\exp{ip_i X(1)}:$ in
(9) for both $Y$ and $J$ -fields:

\begin{eqnarray}
\hat{R}^{out}_{i}= \exp{(\xi_{i}\sum_n \frac {J^{(i)}_{-n}}{n})}\
\exp{(k_{i} \sum_n \frac {Y^{(i)}_{-n}}{n})}\ \exp{ik_{i}
Y^{(i)}_0}\nonumber\\ \widetilde\lambda^{(+)}_i \exp{(-k_{i} \sum_n
\frac {Y^{(i)}_{n}}{n})}\ \exp{(-\xi_{i} \sum_n \frac
{J^{(i)}_{n}}{n})}\;
\end{eqnarray}

\begin{eqnarray}
\hat{R}^{in}_{i+1}= \exp{(-\xi_{i+1}\sum_n \frac
{J^{(i+1)}_{-n}}{n})}\ \exp{(-k_{i+1} \sum_n \frac
{Y^{(i+1)}_{-n}}{n})}\ \exp{(-ik_{i+1} Y^{(i+1)}_0)} \nonumber \\
\lambda^{(-)}_{i+1} \exp{(k_{i+1} \sum_n \frac {Y^{(i+1)}_{n}}{n})}\
\exp{(\xi_{i+1} \sum_n \frac {J^{(i+1)}_{n}}{n})}\
\end{eqnarray}
\begin{eqnarray}
 \sum_i k_i=0
\end{eqnarray}

\begin{eqnarray}
\hat{R}^{(NS)}_{i,i+1}= \exp{(-\zeta_{s}\sum_n \frac
{I^{s}_{-n}}{n})}\ \exp{(-\zeta_{i,i+1}\sum_n \frac {I_{-n}}{n})}\
\exp{(-p_{i,i+1}\sum_n \frac {a_{-n}}{n})}\
\exp{(-ip_{i,i+1}X_0)}\nonumber\\ \Gamma_{i,i+1}
\exp{(p_{i,i+1}\sum_n \frac {a_{n}}{n})}\ \exp{(\zeta_{i,i+1}\sum_n
\frac {I_{n}}{n})}\ \exp{(\zeta_{s}\sum_n \frac {I^{s}_{n}}{n})}\
\end{eqnarray}

   Here we have introduced $\lambda_{\alpha}$ operators to be carrying
quark flavours and quark spin degrees of freedom.

\begin{eqnarray}
\langle0|\widetilde\lambda^{(+)}=0; \lambda^{(-)}|0\rangle=0
\end{eqnarray}

These operators obey simple equations:
\begin{eqnarray*}
&&
\left\{\tilde\lambda_{\alpha}^{(-)},\lambda_{\beta}^{(+)}\right\}\
=\  \delta_{\alpha,\beta}\ ; \quad \tilde\lambda\ =\ \lambda T_0 ,\\
&& T_0=\gamma_0\otimes\tau_2\ ;
\end{eqnarray*}

\begin{eqnarray}
\hat{p}_{i,i+1}=\hat{\beta}^{(i+1)}_{in} \hat{p}_{i+1} +
 \hat{\beta}^{(i)}_{out} \hat{p}_{i}\rightarrow \hat{\beta}^{(i+1)}_{in} k_{i+1} -
\hat{\beta}^{(i)}_{out} k_{i}
\end{eqnarray}

\begin{eqnarray}
(\hat{p}_{i})_{\mu}=\frac{-1}{i}\frac \partial {\partial Y_{0\mu}};
 (\hat{p}_{i+1})_{\mu}=\frac{-1}{i}\frac \partial {\partial Y^{(i+1)}_{0\mu}}
\end{eqnarray}

\begin{eqnarray}
\hat{\zeta}_{i,i+1}=
\hat{\alpha}^{(i+1)}_{in}\hat{\xi}_{i+1}+\hat{\alpha}^{(i)}_{out}\hat{\xi}_{i}
\nonumber \\
\hat{\xi}_{i}= \widetilde\lambda^{(+)}_{i}\xi \lambda^{(-)}_{i};\
\hat{\xi}_{i+1}= \widetilde\lambda^{(+)}_{i+1}\xi
\lambda^{(-)}_{i+1};
\end{eqnarray}
       Here $\xi$ is some universal matrix over quark flavours.

   Let us notice that values $k_i; k_{i+1}; \xi_{i}; \xi_{i+1}; p_{i,i+1}; \zeta_{i,i+1}$ in the vertex operator
$\hat{V}_{i,i+1}(1)$ and in (12),(13),(15) are eigenvalues of the
corresponding operators (18), (17),(19). The quark spinor and
isospinor operators  $\lambda_{\alpha}$  therewith are the
eigenfunctions of the operators $\hat{\xi}_{i}$ and play the same
role as functions
  $\exp{ik_{i} Y^{(i)}_0}$ which are the eigenfunctions of the
operator   $(\hat{p}_{i})_{\mu}=\frac{-1}{i}\frac
\partial {\partial Y_{0\mu}}$.

  So we give some relation between of momenta (charges) which flow
into the basic surface and into edging surfaces.  Namely operators
$\hat{\beta}(\hat{\alpha})$ define  fractions of i-th and (i+1)-th
momenta (charges) for the basic surface. Let us bring definitions
and  constraints for them:

\begin{eqnarray}
\hat{\beta}^{(i+1)}_{in}=\widetilde\lambda^{(+)}_{i+1}\beta_{in}\lambda^{(-)}_{i+1};
\hat{\beta}^{(i)}_{out}=\widetilde\lambda^{(+)}_{i} \beta_{out}\lambda^{(-)}_{i};\nonumber\\
\hat{\alpha}^{(i+1)}_{in}=\widetilde\lambda^{(+)}_{i+1}\alpha_{in}\lambda^{(-)}_{i+1};
\hat{\alpha}^{(i)}_{out}=\widetilde\lambda^{(+)}_{i}\alpha_{out}\lambda^{(-)}_{i};
\end{eqnarray}

Now there are constraints for matrices $\beta;\alpha$:
\begin{eqnarray}
\beta_{in}^2=\beta_{out}^2=\alpha_{in}^2=\alpha_{out}^2=1\\
\beta_{out}=\beta_{in}^+;\alpha_{out}=\alpha_{in}^+
\end{eqnarray}
\begin{eqnarray}
[\beta ,\alpha ] = 0
\end{eqnarray}
  We can propose some simple choice for $\beta_{in},\beta_{out}$:
\begin{eqnarray}
\beta_{in}=\beta=a\gamma^P + b\gamma^C\gamma^P;
\beta_{out}=\beta^+=a\gamma^P-b\gamma^C\gamma^P;
 a=\cosh{\phi};b=\sinh{\phi}
\end{eqnarray}
\begin{eqnarray}
\gamma^C=\gamma^0\gamma^2 ;\gamma^P=\gamma^0;
\end{eqnarray}
Here $\gamma^{\mu}$ are usual  Dirac matrices. The product
$\gamma^C\gamma^P$ corresponds to the CP transformation of spinors
$\lambda$.
  Let us notice that value of $\phi$  defines the fraction of momentum to be
flowing  into two-dimensional surface for the closed string sector
and therefore $\phi \sim 10^{-38}$ as it has discussed above (see
page 3).

\section{ Extended Virasoro superconformal symmetries for composite superstrings}

       The spectrum of spurious states which drop out of physical
 amplitudes and therefore the spectrum of physical composite string states is
 defined by the symmetries of the vertices $\hat{V}_{i,i+1}(1)$.
      Main symmetry of any string model is the superconformal
 symmetry to be defined by the Virasoro operators $G_r$. Certainly
 it requires the operator vertices (11) to have conformal spin j equal
 to be one as for all open string models.
  The operators $G_r$ satisfy standard superconformal algebra:
\begin{eqnarray}
&& \{G_r,G_s\}\ =\ 2L_{r+s}+\frac c3\left(r^2-\frac14\right)
\delta_{r,-s} \\
&& [L_n,L_m]\ =\ (n-m)\ L_{n+m}+\frac c{12}n(n^2-1)\delta_{n,-m}\\
\label{2} && [L_n,G_r]\ =\ \left( \frac n2-r\right) G_{n+r}\ .
\end{eqnarray}
  In the case of the Neveu-Schwarz model we have the generators
$G_r^{NS}$(8) of Virasoro algebra.

    For composite superstring model we consider the set of states
and of superconformal generators for the i-th section between the
$\hat{V}_{i-1,i}$ vertex and $\hat{V}_{i,i+1}$ vertex in (10)(see
Fig.7).

\begin{figure}
\begin{center}
\centerline{\epsfig{figure=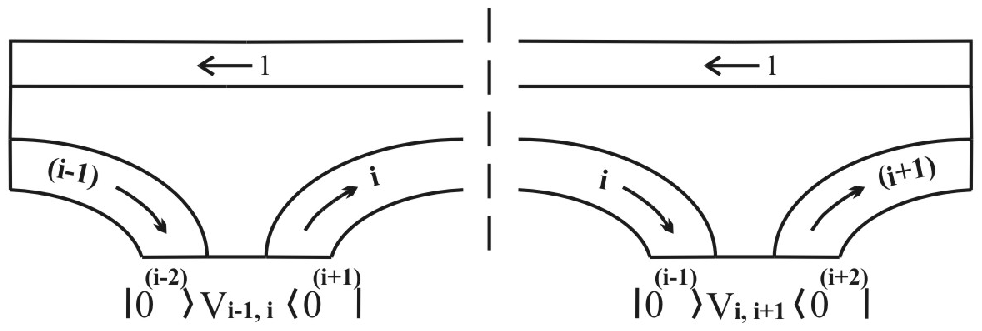,width=0.50\textwidth,clip=}}
Fig.7.
\end{center}
\end{figure}

    Namely we have fields on (i-1), i, (i+1) edging surfaces :\\
        $(Y^{(i-1)},f^{(i-1)});(J^{(i-1)},\Phi^{(i-1)})$;;
        $(Y^{(i)},f^{(i)});(J^{(i)},\Phi^{(i)})$;;$(Y^{(i+1)},f^{(i+1)});(J^{(i+1)},\Phi^{(i+1)})$
        fields

in addition to  fields which are on the basic surface:\\
        $(\partial{X},H)$;$(I,\Theta)$;$(I_s ,\Theta_s)$

  Now superconformal generators $G_r$ can be defined as the
following ones:
\begin{eqnarray}
G_r&=&G^{Lor}_r + G^{Int}_r  \\
G^{Lor}_r&=&\frac1{2\pi}\int\limits_0^{2\pi}d\tau \left(
H^{\mu}\frac d{d\tau} X_\mu +\hat{P}_\nu H^\nu +\hat {p}_1
f^{(1)}+Y^{(i-1)}_\mu f^{(i-1)\mu}+ Y^{(i)}_\mu f^{(i)\mu}+
Y^{(i+1)}_{\mu}f^{(i+1)\mu} \right)e^{-ir\tau} \nonumber
\end{eqnarray}

\begin{eqnarray}
G^{Int}_r&=&\frac1{2\pi}\int\limits_0^{2\pi}d\tau \left(
(I\Theta)+(I_s
\Theta_s)+\xi_1\Phi^{(1)}+(J^{(i-1)}\Phi^{(i-1)})+(J^{(i)}\Phi^{(i)})+(J^{(i+1)}\Phi^{(i+1)})
\right)e^{-ir\tau} \nonumber
\end{eqnarray}

      But unlike the Neveu-Schwarz model this composite string model
      has a new superconformal symmetry which defines by the
      following generators $\widetilde G_r$:
\begin{eqnarray}
&&\widetilde G_r=\widetilde G^{Lor}_r + \widetilde G^{Int}_r \\
&&\widetilde G^{Lor}_r= \frac1{2\pi}\int\limits_0^{2\pi}d\tau
(\left(\partial{X}_\mu \hat{\beta}^{(i-1)}f^{(i-1)\mu}
+\partial{X}_\mu \hat{\beta}^{(i)} f^{(i)\mu}+
\partial{X}_\mu \hat{\beta}^{(i+1)}f^{(i+1)\mu}\right) +\nonumber\\
&&+\left(Y^{(i-1)}_\mu \hat{\beta}^{(i-1)}H^{\mu}+ Y^{(i)}_\mu
\hat{\beta}^{(i)}H^{\mu}+Y^{(i+1)}_{\mu} \hat{\beta}^{(i+1)}H^{\mu}
\right)- \\
&&-(Y^{(i-1)}_\mu\hat{\beta}^{(i-1)}\hat{\beta}^{(i)}f^{(i)\mu}+
Y^{(i)}_\mu\hat{\beta}^{(i)}\hat{\beta}^{(i-1)}f^{(i-1)\mu}+
Y^{(i)}_\mu\hat{\beta}^{(i)}\hat{\beta}^{(i+1)}f^{(i+1)\mu}+\nonumber\\
&&+Y^{(i+1)}_{\mu}\hat{\beta}^{(i+1)}\hat{\beta}^{(i)}
f^{(i)\mu})+\hat {p}_1 f^{(1)})e^{-ir\tau} \nonumber
\end{eqnarray}

\begin{eqnarray}
&& \widetilde G^{Int}_r=\frac1{2\pi}\int\limits_0^{2\pi}d\tau
 (\left(I\hat{\alpha}^{(i-1)}\Phi^{(i-1)}+I\hat{\alpha}^{(i)}\Phi^{(i)}+I\hat{\alpha}^{(i+1)}\Phi^{(i+1)}\right)+\nonumber \\
&&+\left(J^{(i-1)}\hat{\alpha}^{(i-1)}\Theta + J^{(i)}\hat{\alpha}^{(i)}\Theta + J^{(i+1)}\hat{\alpha}^{(i+1)}\Theta\right)-\nonumber\\
&&-(J^{(i-1)}\hat{\alpha}^{(i-1)}\hat{\alpha}^{(i)}\Phi^{(i)}+J^{(i)}\hat{\alpha}^{(i)}\hat{\alpha}^{(i-1)}\Phi^{(i-1)}+
J^{(i+1)}\hat{\alpha}^{(i+1)}\hat{\alpha}^{(i)}\Phi^{(i)}+ \\
&&+ J^{(i)}\hat{\alpha}^{(i)}\hat{\alpha}^{(i+1)}
\Phi^{(i+1)})+I_s\Theta_s +\xi_1\Phi^{(1)})e^{-ir\tau}\nonumber
\end{eqnarray}

   These operators $\widetilde G_r$ have the same commutation
   relations to the operator vertex $\hat{V}_{i,i+1}$ as $G_r$:

\begin{eqnarray}
&&[\widetilde G_r,\hat{V}_{i,i+1}(1)] = [G_r,\hat{V}_{i,i+1}(1)]; \\
&&[\widetilde G_r,\hat{W}_{i,i+1}] = [G_r,\hat{W}_{i,i+1}]
\end{eqnarray}

   Just here we have used the definite relations (17)-(23) for momenta and
charges in the operator vertex $\hat{V}_{i,i+1}$.
  Taking into account the expressions (29)-(32) we
can derive the corresponding commutation relations for $\widetilde
G_r$ and $G_r$:

\begin{eqnarray}
\{G_r,G_s\}\ =\ 2L_{r+s}+\frac c3\left(r^2-\frac14\right)\delta_{r,-s} \\
\{G_r,\widetilde G_s\}\ = 2\widetilde G_{r+s};
\end{eqnarray}

\begin{eqnarray}
\{\widetilde G_r,\widetilde G_s\}\ = 4L_{r+s}-2\widetilde L_{r+s}
 +\frac {2c}3 \left(r^2-\frac14\right)\delta_{r,-s}
\end{eqnarray}

\begin{eqnarray}
[L_n,L_m]=(n-m)L_{n+m}+\frac{c}{12}n(n^2-1)\delta_{n,-m}
\end{eqnarray}
\begin{eqnarray}
[L_n,\widetilde L_m]=(n-m)\widetilde L_{n+m}
\end{eqnarray}
\begin{eqnarray}
[\widetilde L_n,\widetilde L_m]=2(n-m)L_{n+m}-(n-m)\widetilde
L_{n+m} +\frac{c}{6}n(n^2-1)\delta_{n,-m}
\end{eqnarray}
\begin{eqnarray}
[L_n,G_r] =\left( \frac n2-r \right)G_{n+r}
\end{eqnarray}
\begin{eqnarray}
[L_n,\widetilde G_r] =\left( \frac n2-r \right)\widetilde G_{n+r}
\end{eqnarray}
\begin{eqnarray}
[\widetilde L_n,\widetilde G_r] =\left(n-2r \right)G_{n+r}-\left(
\frac n2-r \right)\widetilde G_{n+r}
\end{eqnarray}

  Due to this algebra and equations (33), (34) we are able to prove
as earlier in classical models that both $G_r$ and $\widetilde G_r$
operators generate spurious
  states.
This commutation agebra allows to extract the independent
combinations of operators $G_r$ and $\widetilde G_r$ which define
the spectrum of spurious states.
   We have three sets of this sort :

\begin{eqnarray}
G^I_r= \frac 13(\widetilde G_r+2G_r)
\end{eqnarray}
\begin{eqnarray}
G^{II}_r=(G^{Lor}_r - \widetilde G^{Lor}_r)
\end{eqnarray}
\begin{eqnarray}
G^{III}_r=(G^{Int}_r - \widetilde G^{Int}_r)
\end{eqnarray}

\begin{eqnarray}
\{G^I_r,G^{II}_s\} = \{G^{II}_r,G^{III}_s\} = \{G^I_r,G^{III}_s\} =0
\end{eqnarray}
\begin{eqnarray}
[L^I_n,L^{II}_m] =  [L^{II}_n,L^{III}_m] = [L^I_n,L^{III}_m]=0
\end{eqnarray}

   Now we are able to formulate in i-th section of our amplitude (see Fig.7.) the
extended superconformal constraints for physical states:

\begin{eqnarray}
G^I_r|Phys\rangle=0\ ;\quad L^I_n|Phys\rangle=0\ ; \quad r,n>0\,\
\end{eqnarray}
\begin{eqnarray}
G^{II}_r|Phys\rangle=0\ ;\quad L^{II}_n|Phys\rangle=0\ ; \quad
r,n>0\,\
\end{eqnarray}
\begin{eqnarray}
G^{III}_r|Phys\rangle=0\ ;\quad L^{III}_n|Phys\rangle=0\ ; \quad
r,n>0\,\
\end{eqnarray}
\begin{eqnarray}
\quad L_0|Phys\rangle= \frac 12 |Phys\rangle \ ;
\end{eqnarray}

     Let us notice that our construction of vertices
(12),(13),(15) according to (17)-(19) contains some definite
combinations of fields with Lorentz indices:

\begin{eqnarray}
k_i\widetilde f^{(i)}= k_i (f^{(i)} + \hat{\beta}_{i} H);\nonumber\\
k_i \widetilde Y^{(i)}= k_i (Y^{(i)}+ \hat{\beta}_i \partial X )
\end{eqnarray}

and with internal quantum numbers (with exception of
$I_s;\Theta_s$):

\begin{eqnarray}
\zeta_i \widetilde \Phi^{(i)}=
\zeta_i(\Phi^{(i)}+\hat{\alpha}_i\Theta);\nonumber\\
  \zeta_i\widetilde J^{(i)}= \zeta_i(J^{(i)}+\hat {\alpha}_i I)
\end{eqnarray}

  So in our analysis of the Fock space of states for the i-th
section of the amplitude (10) (see Fig.7.)  we can use these
combinations and orthogonal to them components  $p^t\widetilde
f^{(i)}; p^t \widetilde Y^{(i)} $ with $p^tk_i=0;p^tk_{i-1}=0;$ (on
the left side) and $p^tk_i=0;p^tk_{i+1}=0;$ (on the right side):
\begin{eqnarray}
p^t\widetilde f^{(i)}= p^t (f^{(i)} + \hat{\beta}_{i} H);\nonumber\\
p^t \widetilde Y^{(i)}= p^t(Y^{(i)}+ \hat{\beta}_i \partial X )
\end{eqnarray}
  Let us notice that all combinations (53),(54)and (55) commute
  with operators  $G^{II}_r,G^{III}_r$ and
     $L^{II}_n,L^{III}_n$.

\section{Elimination of states with negative norms and supercurrent
constraints}

        Let us consider the construction of the spectrum generating
algebra for this composite superstring by similar way as in
classical string models [6]. For the given i- th section ( betweeen
$V_{i-1,i}$ and $V_{i,i+1}$) we have fields on i,i-1,i+1 edging
surfaces and fields on the basic surface.
  We are able to build the set of transversal states which are
similar to DDF states for the Neveu-Schwarz model with help of the
operators of type of vertex operators (11) with the conformal spin j
to be equal to one and transversal components of the $\partial X $
and $H$-fields .

     Spurious states for this basis are defined by products of
  operators $G^I_r,G^{II}_r,G^{III}_r$ and
  $L^I_n,L^{II}_n,L^{III}_n$. But only these states are not able to get
  rid of negative norms the spectrum of physical states as it has
  taken place for usual classical string models since the capacity
 of those of them which have negative norms is not enough.
 Powers of $G^{II}_r,G^{III}_r$ and
     $L^{II}_n,L^{III}_n$ do not allowed to do it where as only odd powers
 of $G^I_r,L^I_n$ are not enough. For the Fock space under
 consideration we can obtain states with negative norms not only the
 powers of time components of the $\partial X $
and $H $ fields on the basic surface but as odd powers of components
(if they are time-like ones):$k_{i-1}\widetilde
f^{(i-1)},k_{i}\widetilde f^{(i)},k_{i+1}\widetilde f^{(i+1)}$ and
$k_{i-1} \widetilde Y^{(i-1)},k_i \widetilde Y^{(i)},k_{i+1}
\widetilde Y^{(i+1)}$.  Hence additional conditions for the
composite string model are
 required in order to eliminate all negative norms from the spectrum
 of physical states. There is a simple solution for it. We shall
 require as gauge conditions the supercurrent
  conditions generated  by   $k_i \widetilde {f}^{(i)}$.

  Namely we shall take the following constraints for our vertices:
\begin{eqnarray}
[k_i \widetilde Y^{(i)}_n,\hat{W}_{i,i+1}] =
[\hat{W}_{i,i+1},k_{i+1} \widetilde Y^{(i+1)}_n]=0
\end{eqnarray}
  Then we shall have enough states of negative norms generated by
all gauge constraints $G^I_r,G^{II}_r,G^{III}_r$;
$L^I_n,L^{II}_n,L^{III}_n$ and \\ $k_i \widetilde {f}^{(i)}_r$;$k_i
\widetilde Y^{(i)}_n$;$k_{i-1} \widetilde {f}^{(i-1)}_r$;$k_{i-1}
\widetilde Y^{(i-1)}_n$;$k_{i+1} \widetilde {f}^{(i+1)}_r$;$k_{i+1}
\widetilde Y^{(i+1)}_n$.

   The equations (56) lead  to the conditions:
\begin{eqnarray}
k_i^2 \rightarrow 0;k_{i+1}^2 \rightarrow 0;(k_ik_{i+1})\rightarrow
0;
\end{eqnarray}

  Similarly we obtain
\begin{eqnarray}
[k_{i-1} \widetilde Y^{(i-1)}_n,\hat{W}_{i-1,i}] =
[\hat{W}_{i-1,i},k_{i} \widetilde Y^{(i)}_n]=0 ;
\nonumber\\
k_i^2 \rightarrow 0;k_{i-1}^2 \rightarrow 0;k_ik_{i-1}\rightarrow 0;
\end{eqnarray}

   So our gauge supercurrents are independent and nilpotent ones:
\begin{eqnarray}
[k_i \widetilde Y^{(i)}_n,k_i \widetilde Y^{(i)}_m] =0;
 [k_{i+1} \widetilde Y^{(i+1)}_n,k_i \widetilde Y^{(i)}_m]=0
\end{eqnarray}

    Let us notice that our choice for additional gauge conditions
is appropriate for emission of $\pi$-mesons  (the case of usual
quarks). It gives an explanation for massless $\pi$-mesons and
correct amplitudes for $\pi$-mesons interaction [9]. But other quark
flavours  bring us to gauge supercurrent constrains which contain
not only fields with Lorentz indices $\widetilde Y^{(i)}$ but and
some part of  fields $\widetilde J^{(i)}$ for internal numbers. So
we have to substitute  $k_i \widetilde Y^{(i)}$ supercurrents to
$k^{(LI)}_i \widetilde Y^{(LI)(i)}$ to be some sum of Lorentz fields
and of fields to be carrying internal numbers (their contribution in
usual quarks should be vanishing due to the corresponding operator
$\hat{\xi}$ for $J_n$ in (12),(13),(19)). Then we have the
generalized momentum $k^{(LI)}_i$ instead of $k_i$ and the
conditions $(k^{(LI)}_i)^2=0$ instead of (57) will lead to massive
mesons for other flavours (K-mesons and so on).
 Now we are able to build spectrum generating
algebra (SGA) for our set of states in the same manner as for the
Neveu- Schwarz string model [6].

     We take for the chosen i-th section with the vertex $V_{i-1,i}$ on the left
side  and the vertex $V_{i,i+1}$ on the right side the following
way.

      We shall use the light-like vectors  $k^{(LI)}_i$  from our vertices
  ($(k^{(LI)}_i)^2=0$) and consider a state of the
 generalized momentum  $P^{(gen)}=p_0 + Nk^{(LI)}_i$.

\begin{eqnarray}
\frac {p_0^2}2= -1;(k^{(LI)}_i)^2=0;(k^{(LI)}_i p_0)=1\\
(k^{(LI)}_{i-1})^2=0;(k^{(LI)}_ {i+1}
   )^2=0;(k^{(LI)}_ik^{(LI)}_{i+1})\rightarrow 0;
(k^{(LI)}_ik^{(LI)}_{i-1})\rightarrow 0;
\end{eqnarray}
    Transversal components of $k^{(LI)}_i,p_0$ are vanishing $(p_0)_a
=(k^{(LI)}_i)_a=0$. The generalized mass of this state is given by :
\begin{eqnarray}
\frac {M^{(gen)2}}2=\frac{(p_0+ Nk^{(LI)})^2}2 =-1+N
\end{eqnarray}

    We define the transversal operators of SGA as corresponding
vertex operators. So for components corresponding to $((Y^{(i)}+
\hat{\beta}_i
\partial X )_{a})_n \equiv \widetilde (Y^{(i)}_{a})_n$
 we use  $(S^{(i)}_a)_n$ operators:

\begin{eqnarray}
(S^{(i)}_a)_n = \frac1{2\pi}\int\limits_0^{2\pi}d\tau
V^{(i)}_a(nk^{(LI)}_i,\tau)\\
V^{(i)}_a(nk^{(LI)}_i,\tau)=\exp i{L_0} \tau \left[G_r,\hat
W_a^{(i)}(nk^{(LI)}_i,0) \right]\exp -i{L_0} \tau
\end{eqnarray}

\begin{eqnarray}
&& \hat W_a^{(i)}(k^{(LI)}_i,\tau)= \frac
1{\sqrt{2}}(f^{(i)}_a(\tau) + \hat{\beta}_{i} H_a(\tau)) :\exp
ik^{(LI)}_i(Y^{(LI)(i)}(\tau)+
\hat{\beta}_i \partial X^{(LI)}(\tau)):\equiv \nonumber \\
&& \frac 1{\sqrt{2}} \widetilde f^{(i)}_a(\tau) :\exp{(ik^{(LI)}_i
\widetilde Y^{(LI)(i)}(\tau))}
\end{eqnarray}
Similarly for components corresponding to $(\widetilde Y^{(i-1)})_n$
 we have $(S^{(i-1)}_a)_n$ operators:
\begin{eqnarray}
(S^{(i-1)}_a)_n = \frac1{2\pi}\int\limits_0^{2\pi}d\tau
V^{(i-1)}_a(nk^{(LI)}_i,\tau)\\
V^{(i-1)}_a(nk^{(LI)}_i,\tau)=\exp i{L_0} \tau \left[G_r,\hat
W_a^{(i-1)}(nk^{(LI)}_i,0) \right]\exp -i{L_0} \tau
\end{eqnarray}

\begin{eqnarray}
\hat W_a^{(i-1)}(k^{(LI)}_i,\tau)= \frac 1{\sqrt{2}}
   \widetilde f^{(i-1)}_a(\tau):\exp {(ik^{(LI)}_i \widetilde Y^{(LI)(i)}(\tau))}\:
\end{eqnarray}

For components corresponding to $(\widetilde Y^{(i+1)})_n$ we have
$(S^{(i+1)}_a)_n$ operators:

\begin{eqnarray}
(S^{(i+1)}_a)_n = \frac1{2\pi}\int\limits_0^{2\pi}d\tau
V^{(i+1)}_a(nk^{(LI)}_i,\tau)\\
V^{(i+1)}_a(nk^{(LI)}_i,\tau)=\exp i{L_0} \tau \left[G_r,\hat
W_a^{(i+1)}(nk^{(LI)}_i,0) \right]\exp -i{L_0} \tau
\end{eqnarray}

\begin{eqnarray}
\hat W_a^{(i-1)}(k^{(LI)}_i,\tau)= \frac 1{\sqrt{2}}
   \widetilde f^{(i+1)}_a(\tau):\exp {(ik^{(LI)}_i \widetilde Y^{(LI)(i)}(\tau))}\:
\end{eqnarray}

 Other transversal SGA operators which are  corresponding to
 transfer of internal quark numbers (flavour, chirality) and
 do not enter in $\widetilde Y^{(LI)}$ ; $\widetilde f^{(LI)}$
 (we shall
 mark them as $\Phi',\Theta'$ and $J',I'$) are defined
 by a similar way :

\begin{eqnarray}
\hat W_a^{Int(i)}(k^{(LI)}_i,\tau)= \frac 1{\sqrt{2}}( \widetilde
\Phi'^{(i)} :\exp {(ik^{(LI)}_i \widetilde Y^{(LI)(i)}(\tau))}\:
\end{eqnarray}

\begin{eqnarray}
\hat W_a^{Int(i-1)}(k^{(LI)}_i,\tau)= \frac
1{\sqrt{2}}(\Phi'^{(i-1)}(\tau) + \hat{\alpha}_{i-1} \Theta '(\tau))
:\exp{( ik^{(LI)}_i(Y^{(LI)(i)}(\tau))}:
\end{eqnarray}

\begin{eqnarray}
\hat W_a^{Int(i+1)}(k^{(LI)}_i,\tau)= \frac 1{\sqrt{2}}( \widetilde
\Phi'^{(i+1)} :\exp {(ik^{(LI)}_i \widetilde Y^{(LI)(i)}(\tau))}\:
\end{eqnarray}
      All these SGA operators $(S^{(i)}_a)_n$ have correct gauge and
commutation properties:

\begin{eqnarray}
[\widetilde G_r,(S^{(i)}_a)_n] = [G_r,(S^{(i)}_a)_n]=0;
\end{eqnarray}

\begin{eqnarray}
[k^{(LI)}_i \widetilde
Y^{(LI)(i)},(S^{(i)}_a)_n]=0;[k^{(LI)}_{i-1}\widetilde
Y^{(LI)(i-1)},(S^{(i)}_a)_n]=0
\end{eqnarray}

\begin{eqnarray}
[k^{(LI)}_i \widetilde f^{(LI)(i)},(S^{(i)}_a)_n]=0;[k^{(LI)}_{i-1}
\widetilde f^{(LI)(i-1)},(S^{(i)}_a)_n]=0;
\end{eqnarray}

      As for the Neveu-Schwarz model more complicated constructions  are
used for independent transversal SGA operators $(B^{(i)}_a)_r$
corresponding to
 $(f^{(i)}_r + \hat{\beta}_{i} H_r)\equiv \widetilde f^{(i)}_r $ components:

\begin{eqnarray}
(B^{(i)}_a)_r = \frac1{2\pi}\int\limits_0^{2\pi}d\tau
U^{(i)}_a(rk^{(LI)}_i,\tau) \\
U^{(i)}_a(k^{(LI)}_i,\tau)=\exp {L_0} \tau \left[G_r,\hat
Z_a^{(i)}(k^{(LI)}_i,0) \right]\exp -i{L_0} \tau
\end{eqnarray}
\begin{eqnarray}
&&Z_a^{(i)}(k^{(LI)}_i,\tau )= -\frac 1{\sqrt{2}} \widetilde
f^{(i)}_a(\tau)(k^{(LI)}_i  \widetilde f^{(i)}(\tau))
\\& &(k^{(LI)}_i \widetilde Y^{(LI)(i)}(\tau))^{\frac {-1}2}:\exp {(ik^{(LI)}_i \widetilde Y^{(LI)(i)}(\tau)
)}:\nonumber
\end{eqnarray}

     These SGA operators satisfy necessary constraints:

\begin{eqnarray}
\{G_r,(B^{(i)}_a)_s \}=\{\widetilde G_r,(B^{(i)}_a)_s \}=0;
\end{eqnarray}
\begin{eqnarray}
[k^{(LI)}_i \widetilde
Y^{(LI)(i)},(B^{(i)}_a)_r]=0;[k^{(LI)}_{i-1}\widetilde
Y^{(LI)(i-1)},(B^{(i)}_a)_r]=0
\end{eqnarray}
\begin{eqnarray}
\{k^{(LI)}_i \widetilde f^{(LI)(i)},(B^{(i)}_a)_r \}=0 ;
\{k^{(LI)}_{i-1} \widetilde f^{(LI)(i-1)},(B^{(i)}_a)_r \}=0;
\end{eqnarray}

     Similarly we build transversal SGA operators $(\Theta^{(i)})_r$ which are carrying
internal numbers and correspond to the combinations which do not
enter in $\widetilde Y^{(LI)}$ ; $\widetilde f^{(LI)}$ i.e.
correspond to the $\Phi',\Theta'$ and $J',I'$) supercurrents:
\begin{eqnarray}
(\Theta^{(i)})_r = \frac1{2\pi}\int\limits_0^{2\pi}d\tau
U^{Int(i)}(rk^{(LI)}_i,\tau) \\
U^{Int(i)}(k^{(LI)}_i,\tau)=\exp i{L_0} \tau \left[G_r,\hat
Z^{Int(i)}(k^{(LI)}_i,0) \right]\exp -i{L_0} \tau \
\end{eqnarray}

\begin{eqnarray}
&&Z^{Int(i)}(k^{(LI)}_i,\tau )= -\frac 1{\sqrt{2}}\widetilde
\Phi'^{(i)}(\tau)(k^{(LI)}_i \widetilde f^{(i)}(\tau))
\\& &(k^{(LI)}_i\widetilde Y^{(LI)(i)}(\tau))^{\frac {-1}2}:\exp {(ik^{(LI)}_i(\widetilde Y^{(LI)(i)}(\tau))}\:\nonumber
\end{eqnarray}

   Again we have similar expressions for $(\Theta^{(i-1)})_r$ and
   $(\Theta^{(i+1)})_r$.

         All transversal SGA operators satisfy simple commutation algebra:
\begin{eqnarray}
&&[(S^{(i)}_a)_n,(S^{(j)}_b)_m]=m\delta^{i,j}\delta_{a,b}\delta_{m+n,0}  \\
&&[(S^{(i)}_a)_n,(B^{(j)}_b)_r]=0;
\{(B^{(i)}_a)_r,(B^{(j)}_b)_s\}=\delta^{i,j}\delta_{a,b}\delta_{m+n,0}
\nonumber
\end{eqnarray}     So we can construct similarly to the DDF states transversal
 states $|Phys\rangle$ from powers of the transversal SGA operators:
\begin{eqnarray}
|Phys\rangle =
 \prod {((S^{(i)}_a)_{-n})^{\lambda (a,n)}}...|\Psi_0\rangle
\end{eqnarray}

     These states
will satisfy the following conditions:

\begin{eqnarray}
G_r|Phys\rangle = 0;   L_n|Phys\rangle = 0;    n>0;r>0 \\
 \widetilde G_r\rangle = 0;\widetilde L_n|Phys\rangle = 0; \nonumber \\\nonumber
(\widetilde G^{Lor}_r - G^{Lor}_r)|Phys\rangle = 0; (\widetilde
L_n^{Lor} - L^{Lor}_n)|Phys\rangle = 0;\nonumber \\(k^{(LI)}_i
\widetilde Y^{(LI)(i)})_n|Phys\rangle = 0;(k^{(LI)}_{i-1} \widetilde
Y^{(LI)(i-1)})_n|Phys\rangle = 0; \nonumber \\(k^{(LI)}_i \widetilde
f^{(LI)(i)})_r|Phys\rangle = 0;(k^{(LI)}_{i-1} \widetilde
f^{(LI)(i-1)})_r|Phys\rangle = 0;\nonumber
\end{eqnarray}

    Let us notice that all transversal SGA operators on the left side with
    (i-1)- and (i)- operators
( (i+1)-operators are vanishing there)  can be defined in
(63)-(68),(72),(73),(78)-(80) and in (84)-(86) with replacement of
 all (i)-fields to (i-1)-fields and vice versa of all (i-1)-fields to (i)-fields.
  It is true and for
all transversal SGA operators on the right side with (i+1)- and (i)-
operators  ( (i-1)- operators are vanishing there ). They can be
defined with replacement of all (i)-fields to (i+1)-fields and vice
versa of all (i+1)-fields to (i)-fields. This possibility to
reformulate these sets of states allows to move from states of i-th
section under consideration to states in (i-1)-th section and so on.

   Moving from these DDF type states to arbitrary states  we can
obtain them as usually with help of ordered powers of the conformal
generators $G^I_r,L^I_n;_n;G^{II}_r,L^{II}_n;G^{III}_r,L^{III}_n$
and  of  powers  of  the  supercurrent operators \\$k^{(LI)}_i
\widetilde Y^{(LI)(i)}$, $k^{(LI)}_i \widetilde f^{(LI)(i)}$ acting
on $|Phys\rangle$  states:

\begin{eqnarray}
(G^I_{\frac {-1}2})^{\lambda (1)}(G^I_{\frac {-3}2})^{\lambda
(3)}...(L^I_{-1})^{\mu (1)}(L^I_{-2})^{\mu (2)}...\prod_r
(k^{(LI)}_{i-1} \widetilde f^{(LI)(i-1)}_{-r})^{\gamma (i-1,r)}
\nonumber \\ \prod_n (k^{(LI)}_{i-1} \widetilde
Y^{(LI)(i-1)}_{-n})^{\delta (i-1,n)} \prod_r (k^{(LI)}_{i}
\widetilde f^{(LI)(i)}_{-r})^{\gamma (i,r)}\prod_n (k^{(LI)}_{i}
\widetilde Y^{(LI)(i)}_{-n})^{\delta (i,n)}|Phys\rangle
\end{eqnarray}

     Then we can repeat considerations in the Neveu-Schwarz
model [6] for  the  theorem about absence of ghosts in the spectrum
of physical states in our case for the critical value of the number
of effective dimension  in relation to $G^I_{\frac{3}2}$ and
$G^I_{\frac {1}2}$  operators and taking into account the conditions
(60), (61).

In critical case the operators $G^I_{\frac {1}2}$ and
$G^I_{\frac{3}2}+ 2 (G^I_{\frac {1}2})^3=G'^I_{\frac{3}2}$\\
define null states:
\begin{eqnarray} \{G'^I_{\frac{3}2},G^I_{\frac{-1}2}\}=0
\end{eqnarray}
\begin{eqnarray}
|S_{\frac{-3}2}\rangle = G'^I_{\frac{-3}2}|Phys\rangle;\langle S_{\frac{3}2}|S_{\frac{-1}2}\rangle =0 \\
| S_{\frac{-1}2}\rangle =
G^I_{\frac{-1}2}|Phys\rangle;\langle|S_{\frac{1}2}
|S_{\frac{-1}2}\rangle =0\nonumber
\end{eqnarray}

\begin{eqnarray}
\langle|S_{\frac{3}2} |S_{\frac{-3}2}\rangle =0
\end{eqnarray}
    The critical case corresponds to the condition (93)
 It requires definite values of numbers of fields:
\begin{eqnarray}
d_{crit}= d_1 +\frac{3}2 d_s =15
\end{eqnarray}
 with the condition:
\begin{eqnarray}
L_0=\widetilde L_0=\frac{1}2
\end{eqnarray}

  Here $d_s$ is the number of isotopic scalar two-dimensional fields
$I_s$ on the basic surface which have no partners on edging
surfaces. And $d_1$ is the number of other fields.
    For the critical case we can prove by the same way as in the
Neveu-Schwarz model that the powers indices $\lambda $ and $\mu$ in
(90) are vanishing  if all constraints for physical states are
fullfilled and hence the norms of all physical states are
nonnegative.

    Let us consider the critical value of the effective dimension
    $d_{crit} = d_1 + \frac {3}2 d_s =15$.
Since $d_1 $ is a number of two sets of fields
 ($\widetilde Y^{(i-1)}, J^{(i-1)}$ and $\widetilde Y^{(i)}, \widetilde J^{(i)}$
 on the left or
($\widetilde Y^{(i+1)}, J^{(i+1)}$ and $\widetilde
Y^{(i)},\widetilde J^{(i)}$ on the right) and hence $d_1 $ is an
even number. It gives only
 $d_s=2,6,10$ but the values 6,10 give only three or zero for the
 number of Y-fields that is not enough for our Minkovsky space-time.
So we have only one possibility: $d_s=2$;$d_1 =12$. That means four
fields for all Y-fields i.e. $Y^{(i)}_{\mu},\mu=0,1,2,3$ and two
$J^{(i)}$-fields: $d_1=12=2(4+2)$.
  A natural choice for two supercurrents $\widetilde J^{(i)}$ are
two chiral nonabelian isotopic currents that corresponds two terms
in $G_r$, namely  $(-i)(\Phi^{L}_1\Phi^{L}_2\Phi^{L}_3 +
\Phi^{R}_1\Phi^{R}_2\Phi^{R}_3)\equiv J^{L}_3\Phi^{L}_3 +
J^{R}_3\Phi^{R}_3$ for each of i,i-1,i+1-th  edging surfaces. It
means in our above consideration $J^{L(i)}_n$ and $J^{R(i)}_n$
components of currents with $(J^{L(i)})_0=
\widetilde\lambda^{(+)}_{i}(1+\gamma_5)\xi^L\lambda^{(-)}_{i}$ and
$(J^{R(i)})_0=
  \widetilde\lambda^{(+)}_{i}(1-\gamma_5)\xi^R\lambda^{(-)}_{i}$.

   So  our currents satisfy a nonabelian Kac-Moody algebra for L and R currents:
\begin{eqnarray}
[(J^{L(i)}_a)_n, (J^{L(i)}_b)_m] = i\epsilon_{abc}
(J^{L(i)}_c)_{m+n}+ n\delta_{ab}\delta_{n,-m}
\end{eqnarray}

\begin{eqnarray}
&&[(J^{R(i)}_a)_n, (J^{R(i)}_b)_m] = i\epsilon_{abc}
(J^{R(i)}_c)_{m+n}+ n\delta_{ab}\delta_{n,-m} \\
&&[(J^{L(i)}_a)_n,(J^{R(i)}_b)_m ]=0
\end{eqnarray}

   The rest two $I_s$-fields are abelian isotopic scalar fields for
left chiral and right chiral currents on the basic surface.

  Just nonabelian Kac-Moody currents lead us to the fermion
two-dimensional fields dominance. Namely the number of fermion
fields exceeds the number of boson fields in our composite
superstring model. (Let us notice that we have a supersymmetry only
on the world two-dimensional surface.) This fermion dominance gives
superconvergence for one-loop planar string diagrams (Fig.8.).

\begin{figure}
\begin{center}
\centerline{\epsfig{figure=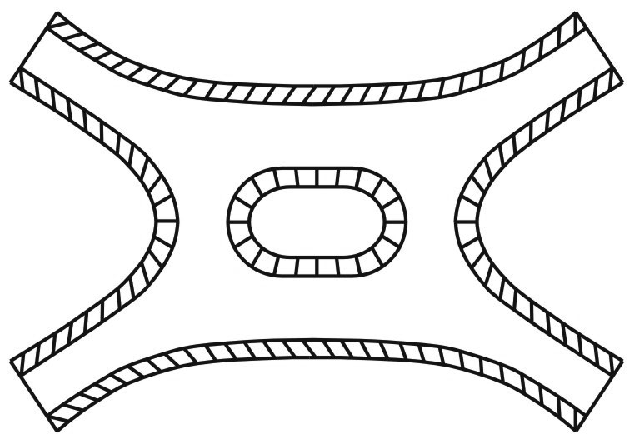,width=0.50\textwidth,clip=}}
Fig.8.
\end{center}
\end{figure}
\section{Conclusion}

    So we have the consistent composite string model for hadron
interactions. It can give the realistic spectrum of physical hadron
states with the leading meson trajectory $\alpha_{\rho}(t)=\frac{1}2
+ \alpha't$. This spectrum is free from states of negative norms.
This model possesses an extended Virasoro superconformal symmetry on
world surfaces (basic and two edging ) and an additional gauge
symmetry generated by nilpotent supercurrents. Just these
supercurrent symmetries lead to the zeroth mass of pions and to the
intercept of $\rho$-trajectory $\alpha_{\rho}(0)$ to be equal to one
half.
 Due to all symmetries it is possible to build the spectrum
 generating algebra of operators for critical case and to prove
 absence of negative norms in the spectrum of physical states
 in this composite string model.

      We have in this approach instead of point-like quarks at
ends of string two-dimensional fields on two edging surfaces which
are carrying quark quantum numbers.

     It is worth to notice that these quark nontrivial quantum
numbers lead to an interesting  possibility of a consistent
description of closed string sector to be arising for nonplanar loop
diagrams in this model.

     In order to describe fermions (baryons) we should move to
the Ramond picture on basic two-dimensional surface . In so doing we
shall hold the relations between edging and basic fields for this
three quark configuration (quantum numbers for two quarks on two
edging surfaces and an Ramond quark is on the basic surface).
     In the following publications we shall consider amplitudes with
different flavours of hadrons , lightest states of a closed string
sector and  one-loop corrections for this composite superstring
model.

      I would like to thank participants of theoretical seminar of PNPI
and HSQCD-2008 participants for attention to this work and useful
discussions of results.

\end{document}